\documentclass[preprint,showpacs,preprintnumbers,amsmath,amssymb]{revtex4}

\usepackage{bm}% bold math

\begin{document}

%\preprint{APS/123-QED}

\title{Surprises in the relativistic free-particle \\ quantization on the light-front}

\author{A.T.Suzuki}
 \altaffiliation[ ]{On sabbatical leave from\\
Permanent address: Instituto de F\'{\i}sica Te\'orica-UNESP,\\
Rua Pamplona 145 -- 01405-900 S\~ao Paulo, SP -- Brazil.}%Lines break automatically or can be forced with \\
\affiliation{Department of Physics, North Carolina State University,
Raleigh, NC  27695-8202}

\author{J.H.O.Sales}
\affiliation{Faculdade de Tecnologia de S\~ao Paulo-DEG,
Pra\c{c}a Coronel Fernando Prestes,
01124-060 S\~ao Paulo, SP}

\author{G.E.R.Zambrano}
\affiliation{Instituto de F\'{\i}sica Te\'orica-UNESP,
Rua Pamplona, 145, 01405-900 S\~ao Paulo, SP}

\date{\today}% It is always \today, today,

\begin{abstract}
We use the light front ``machinery'' to study the behavior of a
relativistic free particle and obtain the quantum commutation
relations from the classical Poisson brackets. We argue that their
usual projection onto the light-front coordinates from the covariant
commutation relations show that there is an inconsistency in the
expected correlation between canonically conjugate variables ``time''
and ``energy''. Moreover we show that this incompatibility originates
from the very definition of the Poisson brackets that is employed and
present a simple remedy to this problem and envisages a profound
physical implication on the whole process of quantization.
\end{abstract}

\pacs{12.39.Ki,14.40.Cs,13.40.Gp}% PACS, the Physics and Astronomy
\maketitle

\section{Introduction}

According to Dirac \cite{dirac49}, it is possible to build forms of
relativistic dynamics for a given system to describe its evolution
from a initial state in any space-time surface whose lengths between
two points lacks a causal connection. As in the non relativistic case,
where the time evolution may be seen as describing a trajectory in the
three-dimensional space, the dynamical evolution of a relativistic
particle can be thought of as the system following a given path, or
trajectory through the hypersurfaces. For example, the hypersurface
$t=0$ defines our three-dimensional space, and is invariant under
translations and rotations. Here any transformation of inertial
reference frames involving ``boosts'' introduces a modification in the
time coordinate, and therefore in the hypersurface at $t=0$. Other
hypersurfaces can be invariant under some kind of ``boost''; for
example, if we define the hypersurface $x^+=t+z$, a ``boost'' in the
$z$-direction does not affect the hypersuface. This hypersurface is
commonly named the null plane, and the coordinate $x^+$ is commonly
referred to as the ``time'' coordinate for the front form (that is,
light-front), since the hypersurface is tangent to the light-cone.

We use the light-front machinery to study the behavior of a relativistic
free particle and obtain the quantum commutation relations from the
classical Poisson brackets. We start off by employing the traditional
Poisson brackets definition for the light-front coordinates and show that
the na\"{i}ve projection from the covariant case is verified. However, we
argue that such an usual projection onto the light-front coordinates
for those brackets from the covariant commutation relations leads
to an incompatible relationship between light-front canonically conjugate
time-energy variables. We argue that this incompatibility originates from
the very definition of the Poisson brackets employed, and present a simple
remedy to this, which not only corrects the right relationship between
light-front canonical variables, but also introduces a very profound
physical modification in the whole process of quantization.

The lay out for our paper is as follows. First we consider the
well-known covariant quantization procedure for a relativistic free
particle as a platform from which we proceed to analyse the issues
raised for the quantization scheme in the light-front
coordinates. Then, we consider the Poisson brackets using light-front
coordinates and the usual definition of Poisson brackets and show that
it agrees with the na\"{i}ve projection onto light-front coordinates
from the covariant Poisson brackets. Next we consider the relevant
incompatibilities and the cure for it. We especially point out that
such an incompatibility is not solely a curious feature of the
traditional light-front quantization, but rather it is responsible, in
this case, for the appearance of the zero mode problem in the
quantization procedure. It does involve therefore a profound physical
implication, since energy and momentum coordinates get entangled and
mixed up within the usual formalism. Then the final section is devoted
to the conclusions where we list five main results that can be drawn
from this work.
\bigskip

\section{Traditional light-front quantization}

The starting point Poisson brackets definition reads,
\begin{equation}
\left\{ A^{\mu },B_{\nu }\right\} =\sum_{\alpha }\frac{\partial A^{\mu }}{
\partial x^{\alpha }}\frac{\partial B_{\nu }}{\partial p_{\alpha }}-\frac{
\partial A^{\mu }}{\partial p_{\alpha }}\frac{\partial B_{\nu }}{\partial
x^{\alpha }},  \label{2.1}
\end{equation}
which for the coordinates $x^{\mu }$ and conjugate momenta $p_{\nu }$, gives
\begin{equation}
\left\{ x^{\mu },p_{\nu }\right\} =\sum_{\alpha }\frac{\partial x^{\mu }}{
\partial x^{\alpha }}\frac{\partial p_{\nu }}{\partial p_{\alpha }}-\frac{
\partial x^{\mu }}{\partial p_{\alpha }}\frac{\partial p_{\nu }}{\partial
x^{\alpha }}=\delta _{\nu }^{\mu }.  \label{2.2}
\end{equation}

Using this basic result in 
\begin{equation}
\left\{ x^{\mu },p^{2}-m^{2}\right\} =\left\{ x^{\mu },p^{2}\right\}
-\left\{ x^{\mu },m^{2}\right\}  \label{2.3}
\end{equation}
we have (observing that the second term on the right hand side yields zero
straightforwardly) 
\begin{eqnarray*}
\left\{ x^{\mu },p^{2}\right\} &=&p_{\nu }\left\{ x^{\mu },p^{\nu }\right\}
+\left\{ x^{\mu },p_{\nu }\right\} p^{\nu } \\
&=&p^{\nu }\left\{ x^{\mu },p_{\nu }\right\} +\left\{ x^{\mu },p_{\nu
}\right\} p^{\nu }\\
&=&2p^{\nu }\left\{ x^{\mu },p_{\nu }\right\} ,
\end{eqnarray*}
which with (\ref{2.2}) gives, 
\begin{equation}
\left\{ x^{\mu },p^{2}\right\} =2p^{\mu }.  \label{2.4}
\end{equation}

This, for the $\mu =0$ component is therefore 
\begin{equation}
\left\{ x^{0},p^{2}\right\} =2p^{0}=2E,  \label{2.5}
\end{equation}
that is, for this particular time component the Poisson bracket
relates to the total energy of the system. For the light-front case,
it is usual to take the component $\mu =+$ as the ``time'' variable,
so that direct projection yields
\begin{equation}
\left\{ x^{+},p^{2}\right\} =2p^{+}.  \label{2.6}
\end{equation}
which means that the light-front ``time'' variable $x^{+}$ relates to
the {\bf momentum} $p^{+}$, an apparent inconsistency between
canonically conjugate variables. In the covariant case we have a
correlation between time and energy in the Poisson brackets, while in
the light front this correlation is clearly lost. It suggests us that
a deeper and more detailed investigation of what is happening with the
Poisson brackets in the light-front is in order.

In the following, we shall work out the Poisson brackets (\ref{2.4})
projected directly onto the light front coordinates, i.e. 
\begin{eqnarray}
\left\{ x^{\mu },p^{+}p^{-}-p^{\perp 2}\right\} ^{\text{lf}} &=&\left\{
x^{\mu },p^{+}p^{-}\right\} -2p^{\perp }\left\{ x^{\mu },p^{\perp }\right\} 
\label{2.7} \\
&=&p^{+}\left\{ x^{\mu },p^{-}\right\} +\left\{ x^{\mu },p^{+}\right\}
p^{-}-2p^{\perp }\left\{ x^{\mu },p^{\perp }\right\} .  \nonumber
\end{eqnarray}
To do this, we need to know the relation (\ref{2.2}) in the light front,
that is, 
\begin{eqnarray*}
\left\{ x^{\mu },p^{-}\right\} ^{\text{lf}} &=&\sum_{\alpha }\frac{\partial
x^{\mu }}{\partial x^{\alpha }}\frac{\partial p^{-}}{\partial p_{\alpha }}-
\frac{\partial x^{\mu }}{\partial p_{\alpha }}\frac{\partial p^{-}}{\partial
x^{\alpha }} \\
&=&\sum_{\alpha }\frac{\partial x^{\mu }}{\partial x^{\alpha }}\frac{
g^{-\beta }\partial p_{\beta }}{\partial p_{\alpha }}-\frac{\partial x^{\mu }
}{\partial p_{\alpha }}\frac{g^{-\beta }\partial p_{\beta }}{\partial
x^{\alpha }} \\
&=&\sum_{\alpha }\delta ^{\mu \alpha }g^{-\beta }\delta _{\beta \alpha }
\text{ ,}
\end{eqnarray*}
so that 
\begin{equation}
\left\{ x^{\mu },p^{-}\right\} ^{\text{lf}}=g^{\mu -}  \label{2.8a}
\end{equation}

In an analogous way, we have
\begin{equation}
\left\{ x^{\mu },p^{+}\right\} ^{\text{lf}}=g^{\mu +}  \label{2.8b}
\end{equation}
and 
\begin{equation}
\left\{ x^{\mu },p^{\perp }\right\} ^{\text{lf}}=g^{\mu \perp }\text{ .}
\label{2.8c}
\end{equation}

Going back to (\ref{2.7}), we get 
\begin{equation}
\left\{ x^{\mu },p^{+}p^{-}-p^{\perp 2}\right\} ^{\text{lf}}=p^{+}g^{\mu
-}+g^{\mu +}p^{-}-2p^{\perp }g^{\mu \perp }\text{.}  \label{2.9}
\end{equation}

In the case of $\mu =+$ component we have 
\[
\left\{ x^{+},p^{2}\right\} ^{\text{lf}}=p^{+}g^{+-}=2p^{+} 
\]
where the metric is defined as 
\begin{equation}
g^{\mu \nu }=\left( 
\begin{array}{rrrr}
0 & 2 & 0 & 0 \\ 
2 & 0 & 0 & 0 \\ 
0 & 0 & -1 & 0 \\ 
0 & 0 & 0 & -1
\end{array}
\right) \text{ \ \ \ \ \ \ \ \ \ \ \ \ \ and \ \ \ \ \ \ }g_{\mu \nu
}=\left( 
\begin{array}{ccrr}
0 & 1/2 & 0 & 0 \\ 
1/2 & 0 & 0 & 0 \\ 
0 & 0 & -1 & 0 \\ 
0 & 0 & 0 & -1
\end{array}
\right) .  \label{2.10}
\end{equation}

This result agrees perfectly with the covariant case (\ref{2.4}), when
projected directly onto the light-front (\ref{2.5}). Note, however,
that the ``time'' variable $x^+$ here is correlating with the
``momentum'' component $p^+$, and not with the ``energy'' component
$p^-$ as in the case of the covariant brackets, where the time $x^0$
correlates with energy $p^0$.

The primary constrainst $\phi_1=p^2-m^2\approx 0$ translated into
light-front variables reads  
\begin{equation}
\phi _{1}=p^{+}p^{-}-p^{\perp 2}-m^{2}\approx 0\text{ ,}  \label{2.11}
\end{equation}

The light front Lagrangian for the free relativistic particle is 
\begin{eqnarray}
{\cal L}_{\text{lf}} &=&-m\sqrt{\dot{x}^{\mu }\dot{
x}_{\mu }}  \nonumber \\
&=&-m\sqrt{\dot{x}^{+}\dot{x}^{-}-\text{ }
\dot{x}^{\perp 2}}\text{.}  \label{2.12}
\end{eqnarray}
from which we can immediately obtain the corresponding canonically
conjugate momentum components, which read
\begin{eqnarray}
p^{-} &=&-m\frac{\dot{x}^{-}}{\sqrt{\dot{x}^{2}}}
\text{, }  \label{2.12a} \\
p^{+} &=&-m\frac{\dot{x}^{+}}{\sqrt{\dot{x}^{2}}}
\text{ \ \ \ \ and} \\
\text{ \ \ }p^{\perp } &=&-m\frac{\dot{x}^{\perp }}{\sqrt{
\dot{x}^{2}}}
\end{eqnarray}

The Hamiltonian is therefore
\begin{eqnarray}
H_{c}^{\text{lf}} &=&p\dot{x}-{\cal L}_{\text{lf}}  \label{2.14}
\\
&=&-m\frac{\dot{x}^{+}}{\sqrt{\dot{x}^{2}}}
\dot{x}^{-}+m\frac{\dot{x}^{\perp 2}}{\sqrt{
\dot{x}^{2}}}-\left( -m\sqrt{\dot{x}^{+}
\dot{x}^{-}-\text{ }\dot{x}^{\perp 2}}\right) =0.  \nonumber
\end{eqnarray}

Since the canonical Hamiltonian vanishes, this gives us a hint that we
should work it out with
\[
\widetilde{H}^{\text{lf}}=\lambda \left( p^{+}p^{-}-p^{\perp 2}-m^{2}\right), 
\]
where the $\lambda$ is a ``time''-independent parameter (the so-called
multiplier).

Since constraints are by their very nature non-dynamical equations,
they do not evolve in ``time'', and therefore satisfy the following
Poisson brackets:
\begin{eqnarray*}
\dot{\phi }_{1} &=&\left\{ \phi _{1},\widetilde{H}^{\text{lf}
}\right\} \\
&=&\left\{ p^{+}p^{-}-p^{\perp 2}-m^{2},\lambda \right\} \left(
p^{+}p^{-}-p^{\perp 2}-m^{2}\right)
+\lambda \left\{ p^{+}p^{-}-p^{\perp 2}-m^{2},\;p^{+}p^{-}-p^{\perp
2}-m^{2}\right\} \\
&\approx &0\text{ .}
\end{eqnarray*}

This means that there are no constraints of the secondary class and we
are unable to determine the multiplier $\lambda $. The existence of a
constraint of the primary class means that the theory is invariant
under reparametrizations of the type $s \rightarrow s'=s'(x^+)$, so we
can choose the natural reparametrization as $x^+=s(x^+)$, which
defines a new constraint in the light front
\begin{equation}
\phi _{2}=x^{+}-s\approx 0\text{ ,}  \label{2.15}
\end{equation}
so that 
\begin{eqnarray*}
\left\{ \phi _{1},\phi _{2}\right\} &=&\left\{ p^{+}p^{-}-p^{\perp
2}-m^{2},\;x^{+}-s\right\}  \\
&=&\left\{ p^{+}p^{-},\;x^{+}-s\right\} +\left\{ -p^{\perp 2},\;x^{+}-s\right\}
+\left\{ -m^{2},\;x^{+}-s\right\}  \\
&=&p^{+}\left\{ p^{-},\;x^{+}\right\} +\left\{ p^{+},\;x^{+}\right\}
p^{-}-2\left\{ p^{\perp },\;x^{+}\right\}  \\
&=&-p^{+}g^{-+}+g^{++}p^{-}-2g^{\perp +}
\end{eqnarray*}
or, 
\begin{equation}
\left\{ \phi _{1},\phi _{2}\right\} =-2p^{+}.  \label{2.16}
\end{equation}

Comparing with the covariant case, this Poisson brackets should have
resulted proportional to the energy, but once again we perceive that
there is an inconsistency here, for instead of the ``energy'' $p^{-}$
we get the {\em momentum} $p^{+}$. Would this inconsistency be
harmless or would this bring about a more serious problem in the
light-front dynamics of a free relativistic particle? The answer turns
out to be as unexpected and surprising as it is: it leads us to the
old light-front zero mode problem. Let us see how.

With the results so far obtained, we can go on to constructing the
Hamiltonian through
\[
H^{\text{lf}}=\lambda _{1}\phi _{1}+\lambda _{2}\phi _{2}\;\approx\; 0 
\]
which, by the condition of non evolution in time of constraints, 
\[
\dot{\phi }_{1}^{\text{lf}}\;\approx \;0\;\approx \;\lambda _{1}\left\{
\phi _{1}^{\text{lf}},\phi _{1}^{\text{lf}}\right\} +\lambda _{2}\left\{
\phi _{1}^{\text{lf}},\phi _{2}^{\text{lf}}\right\} +\frac{\partial \phi _{1}
}{\partial s} 
\]
from which we have 
\[
\lambda _{2}\approx 0\text{ ,} 
\]
and for $\dot{\phi }_{2}$
\begin{eqnarray*}
\dot{\phi }_{2}^{\text{lf}} &\;\approx \;&0\approx \lambda
_{1}\left\{ \phi _{2}^{\text{lf}},\phi _{1}^{\text{lf}}\right\} +\lambda
_{2}\left\{ \phi _{2}^{\text{lf}},\phi _{2}^{\text{lf}}\right\} +\frac{
\partial \phi _{2}}{\partial s} \\
&=&2p^{+}\lambda _{1}-1
\end{eqnarray*}
from which 
\[
\lambda _{1}=\frac{1}{2p^{+}}\text{ ,} 
\]
which yields 
\begin{equation}
H=\frac{1}{2p^{+}}\left( p^{+}p^{-}-p^{\perp 2}-m^{2}\right).  \label{2.17}
\end{equation}

In order to obtain the Dirac brackets, let us construct the matrix ${\cal M}$, 
\[
{\cal M}=\left( 
\begin{array}{cc}
{\cal M}_{11} & {\cal M}_{12} \\ 
{\cal M}_{21} & {\cal M}_{22}
\end{array}
\right) 
\]
where 
\[
{\cal M}_{11}=\left\{ \phi _{1},\phi _{1}\right\} =0\;\;\text{ , \
}\;\;{\cal M}_{21}=\left\{ \phi _{2},\phi _{1}\right\} =2p^{+}
\]
and 
\[
{\cal M}_{12}=\left\{ \phi _{1},\phi _{2}\right\} =-2p^{+}\text{ and \ }
{\cal M}_{22}=\left\{ \phi _{2},\phi _{2}\right\} =0 
\]
so that 
\begin{equation}
{\cal M}=\left( 
\begin{array}{cc}
0 & -2p^{+} \\ 
2p^{+} & 0
\end{array}
\right) \text{ \ and \ }{\cal M}^{-1}=\left( 
\begin{array}{cc}
0 & \displaystyle\frac{1}{2p^{+}} \\ 
-\displaystyle\frac{1}{2p^{+}} & 0
\end{array}
\right)  \label{2.18}
\end{equation}

The Dirac brackets then is given by 
\[
\left\{ x^{\mu },p^{\nu }\right\} _{\text{D}}^{\text{lf}}=g^{\mu \nu
}-\left\{ x^{\mu },\phi _{1}\right\} {\cal M}_{12}^{-1}\left\{ \phi _{2},p^{\nu
}\right\} -\left\{ x^{\mu },\phi _{2}\right\} {\cal M}_{21}^{-1}\left\{ \phi
_{1},p^{\nu }\right\} . 
\]

The second term in the right hand side of the above is 
\begin{eqnarray}
\left\{ x^{\mu },\phi _{1}\right\} {\cal M}_{12}^{-1}\left\{ \phi _{2},p^{\nu
}\right\} &=&\left\{ x^{\mu },p^{+}p^{-}-p^{\perp 2}-m^{2}\right\} \frac{1}{
2p^{+}}\left\{ x^{+},p^{\nu }\right\}   \nonumber \\
&=&\left( p^{+}g^{\mu -}+g^{\mu +}p^{-}-2p^{\perp }g^{\mu \perp }\right) 
\frac{1}{2p^{+}}g^{+\nu }  \label{2.19}
\end{eqnarray}
while the last term is 
\begin{equation}
\left\{ x^{\mu },\phi _{2}\right\} {\cal M}_{21}^{-1}\left\{ \phi _{1},p^{\nu
}\right\} =\left\{ x^{\mu },x^{+}\right\} \frac{1}{\left( -2p^{+}\right) }
\left\{ p^{+}p^{-}-p^{\perp 2}-m^{2},p^{\nu }\right\} =0  \label{2.20}
\end{equation}

From (\ref{2.19}) and (\ref{2.20}), we get 
\begin{equation}
\left\{ x^{\mu },p^{\nu }\right\} _{\text{D}}^{\text{lf}}=g^{\mu \nu
}-\left( p^{+}g^{\mu -}+g^{\mu +}p^{-}-2p^{\perp }g^{\mu \perp }\right) 
\frac{1}{2p^{+}}g^{+\nu }  \label{2.21}
\end{equation}

Quantization can now be carried out by taking 
\begin{equation}
\left[ x^{\mu },p^{\nu }\right] _{\text{D}}^{\text{lf}}=i\left[ g^{\mu \nu
}-\left( p^{+}g^{\mu -}+g^{\mu +}p^{-}-2p^{\perp }g^{\mu \perp }\right) 
\frac{1}{2p^{+}}g^{+\nu }\right]  \label{2.22}
\end{equation}

This result clearly shows us that we have here the built-in problem of
zero modes in the light-front.

\section{Inconsistencies in the traditional light-front quantization}

From the very beginning of the description of elementary particles
through quantum mechanical wave functions, we have the following
expansion in terms of plane waves: 
\begin{equation}
\Psi (x)\propto {\rm e}^{ip\cdot x}  \label{2.23}
\end{equation}

The argument for the exponential contains a dot product between two
four vectors , namely, position and momentum, i.e.,
\begin{equation}
p\cdot x=p^{\mu }x_{\mu }=p^{0}x_{0}\;-{\bf p}\cdot {\bf x}  \label{2.24}
\end{equation}
so that the zeroth component piece correlates time and energy as canonically
conjugate variables.

In the light front coordinates, this yields
\begin{eqnarray}
p\cdot x &=&p^{\mu }x_{\mu }=p^{+}x_{+}+p^{-}x_{-}+p^\perp \cdot x_\perp  \label{2.25a} \\
&=&\frac{1}{2}p^{-}x^{+}+\frac{1}{2}p^{+}x^{-}-p^{\perp }\cdot x^{\perp}  \label{2.25b}
\end{eqnarray}
so that here again, we have a consistent correlation between the light-front
``time'' $x^{+}$ and the light-front ``energy'' $p^{-}$.

However, as we have pointed out earlier --- compare (\ref{2.5}) and
(\ref{2.6}) --- the Poisson brackets in the covariant case and its
direct projection onto light-front coordinates show us that such a
projection introduces an inconsistency in the original correlation
between canonically conjugate time and energy variables. Moreover, if we take the above result (\ref{2.21}) and compare it to the covariant case, namely,
\begin{equation}
\{x^\mu,\;p_\nu\}_{\rm D} = \delta^\mu_\nu-\delta^0_\nu\,\frac{p^\mu}{p^0}
\end{equation}
we see that again the Dirac brackets in the light-front coordinates
introduces a violation in the canonically conjugate time-energy
variables. Note that in the second term of the brackets, the term
proportional to the covariant energy $(p^0)^{-1}$ corresponds in the
light-front projection to the {\em momentum} $(p^+)^{-1}$ instead of
the expected ``energy'' $(p^-)^{-1}$. This contradicts the observation
made soon after equation (\ref{2.25b}) and arises from the fact that
in the covariant case, $p^0\equiv p_0$, while in the light-front
coordinates, $p^-\neq p_-$, but $p^-\equiv p_+$ (cf. (\ref{2.25a}) and
(\ref{2.25b})). These relations motivate us to seek a possible
solution to this problem in the very definition of the Poisson
brackets, namely, to take it according to the classical definition
with all space-time components in the contravariant notation (all
covariant notation is fine as well, since classically, they are
equivalent).

\section{Poisson brackets in the light-front}

Dirac emphasized in his works \cite{dirac50,dirac51} that the problem
of finding a new dynamical system reduced to that of finding a new
solution to the Poisson brackets. Therefore, we will make a close
inspection into this and make a careful investigation on the
definition of the Poisson brackets and its physical implications.

To do that, let us first of all see how the Lorentz transformation
looks like in the light-front coordinates. It is well known that a
Lorentz transformation from a given inertial frame of reference $S$ to
another frame of reference $S^{\prime}$ in the ($x^3\equiv
z$)-direction is given by (we take for the speed of light $c=1$):
\begin{eqnarray}
x^{\prime\,3}&=&x^{3}\cosh \eta -x^{0}\sinh \eta   \label{lor1}\\
x^{\prime\,0}&=&x^{0}\cosh \eta -x^{3}\sinh \eta   \label{lor2}
\end{eqnarray}
where
\[
\sinh \eta =\frac{\beta }{\sqrt{1-\beta^2 }}\text{ \ \ and \ \ }\beta =v
\text{ .}
\]

We have omitted the transverse coordinates $x^1$ and $x^2$ which are
not affected by the Lorentz boost in the $x^3$-direction. We point out
that this usual Lorentz transformation in the Minkowskian
four-dimensional space-time mixes up space coordinates $x^3$ with time
coordinates $x^0$.

Dirac observed in 1949 that it may be more convenient to use the light-cone
variables
\[
x^{+}=x^{0}+x^{3}
\]
\[
x^{-}=x^{0}-x^{3}
\]

In terms of theses variables, the Lorentz transformation, (\ref{lor1})
and (\ref{lor2}), becomes
\begin{eqnarray*}
x^{\prime\, +} &=&x^{\prime\, 0}+x^{\prime\, 3} \\
&=&x^{0}\left( \cosh \eta -\sinh \eta \right) +x^{3}\left( \cosh \eta -\sinh
\eta \right) 
\end{eqnarray*}
so that
\begin{equation}
x^{\prime\, +}=e^{-\eta }x^{+} \label{lf11} 
\end{equation}
and in a similar manner we have
\begin{equation}
x^{\prime\, -}=e^{\eta }x^{-}  \label{lf12}
\end{equation}

Observe that now, $x^{+}$ and $x^{-}$ do not become linearly mixed up
under this transformation. Therefore, (\ref{lf11}) and (\ref{lf12})
show us that Lorentz transformation in the light front coordinates
becomes simply a {\em scale} transformation where ``time'' $x^+$ does
not mix up with ``space'' $x^-$.

The above result shows us that now, the invarinat scalar in this
two-dimensional plane $(+,-)$ is the {\em bilinear} $x^+x^-$ since
$x^{\prime\, +}x^{\prime\, -}=x^+x^-$ and not $s^2=(x^+)^2+(x^-)^2$ as
it would be in the usual two-dimensional Euclidean plane
$\mathbb{R}^2$. However, this bilinear can be written down as:
\begin{eqnarray}
s^2&=&x^+x^-=\frac{1}{2}x^+x^-+\frac{1}{2}x^-x^+\nonumber \\
&=&g_{+-}x^+x^-+g_{-+}x^-x^+
\end{eqnarray}
where we have used the appropriate components of the light-front
metric tensor (\ref{2.10}). Thus, the two-dimensional plane
$(+,-)$, is indeed a topological space, or a Hilbert vector space,
whose metric has off-diagonal matrix elements $1/2$. Its character is
Riemannian and Euclidean, though.

Moreover, the $(i,j), i,j=1,2$ sector is clearly a pseudo-Euclidean plane.

Therefore, if we restrict ourselves to the superscript notation to
identify the elements of the light-front components, they live in
sectorized topological spaces of the Euclidean character, that is,
{\em no distinction} between contravariant and covariant
notations. (Similar analysis as done above can be carried out with all
light-front subscript notation, of course.)

\section{Light-front quantization without inconsistencies} 

Considering that in the light-front framework the four-dimensional
space-time is sectorized into two two-dimensional Euclidean subspaces
and the Poincar\'e group as a direct sum of two two-dimensional
orthogonal subgroups, where coordinates $(+,-)$ label one of the
two-dimensional spaces and $(1,2)$ the other one, then we have that
all the indices can be treated as non-relativistic (i.e. no distinction
between covariant and contravariant ones). Therefore, the Poisson brackets
(\ref {2.1}) reads
\begin{equation}
\left\{ A^{\mu },B^{\nu }\right\} =\sum_{\alpha }\frac{\partial A^{\mu }}{%
\partial x^{\alpha }}\frac{\partial B^{\nu }}{\partial p^{\alpha }}-\frac{%
\partial A^{\mu }}{\partial p^{\alpha }}\frac{\partial B^{\nu }}{\partial
x^{\alpha }}  \label{4.1}
\end{equation}
and for the fundamental one between canonically conjugate variables
$x$ and $p$, we have
\begin{equation}
\left\{ x^{\mu },p^{\nu }\right\} =\delta ^{\mu \nu
}\,,\:\:\:\mbox{with}\:\:\:\mu ,\nu =+,-,\perp \label{4.2}
\end{equation}

So, for a relativistic free particle in the light-front coordinates we
have the following Poisson brackets,
\begin{equation}
\left\{ x^{\mu },\,p^{+}p^{-}-p^{\perp 2}-m^{2}\right\} ^{\text{lf}
}=p^{+}\delta ^{\mu -}+\delta ^{\mu +}p^{-}-2p^{\perp }\delta ^{\mu \perp }
\text{,}  \label{4.3}
\end{equation}
which for the ($\mu =+$)-component gives
\begin{equation}
\left\{ x^{+},\,p^{+}p^{-}-p^{\perp 2}-m^{2}\right\} ^{\text{lf}}=p^{-}
\text{.}  \label{4.4}
\end{equation}

This implies that in the previous constraint evaluation
\begin{equation}
\left\{ \phi _{1},\,\phi _{2}\right\} =-p^{-}  \label{4.5}
\end{equation}
and therefore
\[
\dot{\phi }_{1}^{\text{lf}}\approx 0\approx \lambda _{1}\left\{
\phi _{1}^{\text{lf}},\phi _{1}^{\text{lf}}\right\} +\lambda _{2}\left\{
\phi _{1}^{\text{lf}},\phi _{2}^{\text{lf}}\right\} +\frac{\partial \phi _{1}%
}{\partial s} 
\]
giving
\[
\lambda _{2}\approx 0\,,
\]
 whereas for $\dot{\phi }_{2}$ we have
\begin{eqnarray*}
\dot{\phi }_{2}^{\text{lf}} &\approx &0\approx \lambda
_{1}\left\{ \phi _{2}^{\text{lf}},\phi _{1}^{\text{lf}}\right\} +\lambda
_{2}\left\{ \phi _{2}^{\text{lf}},\phi _{2}^{\text{lf}}\right\} +\frac{%
\partial \phi _{2}}{\partial s} \\
&=&p^{-}\lambda _{1}-1
\end{eqnarray*}
leading to
\[
\lambda _{1}=\frac{1}{p^{-}}\text{ ,} 
\]
from which we get the Hamiltonian 
\begin{equation}
H=\frac{1}{p^{-}}\left( p^{+}p^{-}-p^{\perp 2}-m^{2}\right).  \label{4.6}
\end{equation}

The matrix ${\cal M}$ turns out to be now
\begin{equation}
{\cal M}=\left( 
\begin{array}{cc}
0 & -p^{-} \\ 
p^{-} & 0
\end{array}
\right) \text{ \ \ \ and  \ \ \ }{\cal M}^{-1}=\left( 
\begin{array}{cc}
0 & \displaystyle\frac{1}{p^{-}} \\ 
-\displaystyle\frac{1}{p^{-}} & 0
\end{array}
\right).  \label{4.7}
\end{equation}

Therefore, the Dirac brackets becomes
\begin{equation}
\left\{ x^{\mu },\,p^{\nu }\right\} _{\text{D}}^{\text{lf}}=\delta ^{\mu \nu
}-\left( p^{+}\delta ^{\mu -}+\delta ^{\mu +}p^{-}-2p^{\perp }\delta ^{\mu
\perp }\right) \frac{1}{p^{-}}\delta ^{+\nu }  \label{4.8}
\end{equation}

Quantization now is achieved by building upon the commutator
\[
\left[ x^{\mu },\,p^{\nu }\right] _{\text{D}}^{\text{lf}}=i\left( \delta ^{\mu
\nu }-\left( p^{+}\delta ^{\mu -}+\delta ^{\mu +}p^{-}-2p^{\perp }\delta
^{\mu \perp }\right) \frac{1}{p^{-}}\delta ^{+\nu }\right) 
\]

Note the consistency of this Dirac commutator in the light-front as
compared to the covariant commutator. Both on the right-hand-side
appear with a term inversely proportional to the energy of the
system. The zero mode present in the traditional light-front
quantization is gone.

\section{Conclusions}

With this unpretentious exercise on the relativistic free-particle
quantization in the light-front we achieved what we deem to be a very
profound physical significance in the whole process of light-front
quantization and light-front framework for the description of
fundamental interactions. Summarizing them we can say that:

1) There is a satisfactory resolution of the inconsistencies present
in the traditional light-front quantization, where the correlation
between canonically conjugate time-energy variables is violated, and
also beset with zero modes.

2) The energy-momentum relation, $k^- \propto (k^+)^{-1}$, which in the
   usual treatment is linear, indeed is not or cannot be considered as
   such, but probably as a {\em bilinear} in the form of $k^+k^-$;

3) The four-dimensional Minkowski space-time is broken down into two
   sectorized Euclidean two-dimensional subspaces, and the Lorentz
   transformation in the light-front is Euclidean-like; they become in
   fact a scale transformation;

4) Since the light-front formulation has these two distinct,
   characteristic sectors, it is ideal, or at least more suitable, for
   studying massless gauge fields, whose intrinsic two-component
   transverse degrees of freedom are the physical components;

5) Probably, the more serious issues raised by the zero mode problems
   in quantum field theory in the light-front cannot be addressed
   satisfactorily unless the cited bilinear term $k^+k^-$ with the
   constraint $k^+k^-=k^{\perp 2}+m^2$ be satisfactorily dealt with
   with some convenient mathematical tool yet to be envisaged. One can
   easily check that if we treat the two variables as linearly
   independent as in the traditional light-front quantization, i.e.,
   $k^+\propto (k^-)^{-1}$, the four-dimensional momentum integration
   measure will still produce the elusive and provocative zero mode problem. 
 
\vspace{1cm}

{\bf Acknowledgments:} A.T.Suzuki thanks the kind hospitality of
Physics Department, North Carolina Staste University and gratefully
acknowledges partial support from CNPq (Bras\'{\i}lia) in the earlier stages
of this work, now superseded by a grant from CAPES (Bras\'{\i}lia),
J.H.O.Sales thanks the hospitality of Instituto de F\'{i}sica
Te\'{o}rica/UNESP, where part of this work has been done, and 
G.E.R.Zambrano acknowledges total support from CAPES (Bras\'{\i}lia).

\end{document}